\documentclass[aps, prd,twocolumn,preprintnumbers,nofootinbib,tightenlines,superscriptaddress]{revtex4-2}
\usepackage[utf8]{inputenc}
\usepackage{graphicx,amsmath,amsfonts,amssymb,aas_macros,slashed}
\usepackage{bbold}
\usepackage{datetime}
\usepackage{numprint}
\usepackage{enumitem}
\usepackage{booktabs}
\usepackage{numprint}
 \usepackage{mathrsfs}
 \usepackage{verbatim}

\usepackage{tikz}
\usepackage[errorstop]{feynmp}
\usepackage[compat=1.1.0]{tikz-feynman} 

\tikzset{graviton/.style={decorate, decoration={snake, amplitude=.4mm, segment length=1.5mm, pre length=.5mm, post length=.5mm}, double}}

\usepackage{graphicx,color,amsmath}
\usepackage{hyperref}

\usepackage[normalem]{ulem}
\usepackage{cancel}

\newif\ifmarkup
\markupfalse      

\ifmarkup
  \newcommand{\del}[1]{{\color{magenta}\sout{#1}}}
  
\else
  \newcommand{\del}[1]{\unskip}   
 
\fi

\allowdisplaybreaks

\def\d{d}

\def\L{\mathcal{L}}

\def\vec{\mathbf}
\def\ii{\mathrm{i}}
\def\e{\mathrm{e}}

\def\M{\mathcal{M}}

\def\vecp{\vec{p}}
\def\vecq{\vec{q}}

\let\oldhat\hat
 \renewcommand{\hat}[1]{\bm\oldhat{\mathbf{#1}}}

\usepackage{pgfplots}
\usepackage{bm}
\usepackage{mathtools}

\begin{document}
\preprint{UWThPh 2025-19}

\title{Soft Gravitons, Hard Truths: Infrared Safety of Particle Processes in a Gravitational-Wave Background}

\author{Wen-Yuan Ai}
\email{wenyuan.ai@oeaw.ac.at}
\affiliation{State Key Laboratory of Dark Matter Physics,\\
Tsung-Dao Lee Institute and School of Physics and Astronomy,\\ Shanghai Jiao Tong University, Shanghai 201210, China}
\affiliation{Key Laboratory for Particle Astrophysics and Cosmology (MOE), and Shanghai Key Laboratory for Particle Physics and Cosmology, Shanghai Jiao Tong University, Shanghai 201210, China}
\affiliation{
Marietta Blau Institute for Particle Physics, Austrian Academy of Sciences,\\ Dominikanerbastei 16, A-1010 Vienna, Austria}

\author{Sebastian A. R. Ellis}
\email{sebastian.ellis@kcl.ac.uk}
\affiliation{King’s College London, Strand, London, WC2R 2LS, United Kingdom}
\affiliation{Departement de Physique Theorique, Universite de Geneve, 24 quai Ernest Ansermet, 1211 Gen`eve 4, Switzerland}

\author{Josef Pradler}
\email{josef.pradler@oeaw.ac.at}
\affiliation{
Marietta Blau Institute for Particle Physics, Austrian Academy of Sciences,\\ Dominikanerbastei 16, A-1010 Vienna, Austria}
\affiliation{University of Vienna, Faculty of Physics, Boltzmanngasse 5, A-1090 Vienna, Austria}

\begin{abstract}
Gravitational waves are thought to propagate unattenuated through matter due to a cancellation between graviton absorption and stimulated emission inferred from leading-order soft-graviton arguments. We revisit this reasoning and show that it fails for the converse problem: the effect of a gravitational-wave background on matter. 
At leading order, real graviton emission \emph{and} absorption appear to enhance decay rates of unstable particles.
By extending the soft-graviton framework describing real and virtual processes in a gravitational wave background, and resumming them to all orders, we show that inclusive decay rates remain essentially unchanged. Bose-enhanced emission and absorption in graviton spectra persist but compensate each other to all orders; the mutual transparency between matter and gravitational radiation hence follows from infrared safety.
\end{abstract}

\maketitle

\section{Introduction} 
In one of his final papers, Weinberg, with Flauger, asked whether gravitational waves (GWs) could be attenuated through interactions with matter~\cite{Flauger:2019cam}, a question also posed and addressed earlier in~\cite{lightman1975problem,Baym:2017xvh}. Given the weakness of gravity, at first, such an effect seems implausible. Yet, following Weinberg's characteristic insistence on first-principles calculation~\cite{Burgess:2025glf}, they were motivated by the observation that the graviton absorption rate grows steeply toward low frequencies, scaling as $d\Gamma_{\rm abs}\propto \nu^{-3}$.
If this were the whole story, the attenuation length of $\mathrm{nHz}$ gravitational waves would be shorter than a megaparsec. They concluded, however, that the absorption is nearly cancelled by stimulated emission of comparable strength, rendering the Universe transparent to long-wavelength gravitational radiation---an expectation consistent with the possible detection of a stochastic GW background with pulsar-timing arrays~\cite{NANOGrav:2023gor,EPTA:2023fyk,Reardon:2023gzh,Xu:2023wog}.

Motivated by the same spirit, in this \emph{Letter}, we extend this program to ask the converse question:
what influence can a GW background exert on matter itself?
 We are motivated by the observation that for unstable particles~$\phi$, decaying as $\phi\to Y$, both graviton absorption \emph{and} emission open additional decay channels,
\begin{align}
\text{vacuum} &: \quad \phi \to Y, \label{eq:vacuumprocess}\\
\text{absorption} &: \quad \phi + g_{\rm GW} \to Y, \label{eq:absorptionprocess}\\
\text{stimulated emission} &: \quad \phi|_{\rm GW} \to Y + g. \label{eq:emissionprocess}
\end{align}
where the subscript GW denotes the GW background and $g$ is a graviton.
In contrast to~\cite{Flauger:2019cam}, stimulated emission does not a priori provide a reverse channel, that, from the perspective of the decaying~$\phi$ particle, could cancel potentially large corrections to the vacuum decay rates.

The universality of gravity allows for a wide variety of candidate particles~$\phi$: dark matter (DM) with a vacuum lifetime exceeding the age of the Universe, Standard Model neutrino mass eigenstates decaying radiatively, neutrons expelled from neutron star mergers fueling $r$-process nucleosynthesis, or, in the presence of primordial GW backgrounds, neutrons participating in light-element formation in the early Universe. 
As we will demonstrate, a straightforward application of the soft-graviton theorem in a Boltzmann framework, as employed in~\cite{Flauger:2019cam} for supermassive black hole (SMBH) background frequencies, suggests large corrections to the $\phi$ lifetime.

We show that this inference is misguided. The key observation is that the infrared-sensitive rates associated with~\eqref{eq:absorptionprocess} and~\eqref{eq:emissionprocess} become enhanced in a region where leading-order real emission processes are substantially corrected by \emph{in-medium} virtual graviton exchange. A similar observation was already made by Czarnecki, Kamionkowski, Lee, and Melnikov~\cite{Czarnecki:2011mr} in quantum electrodynamics (QED), who studied finite-temperature radiative corrections to charged-particle decay in the early Universe. There, the inclusion of photon emission, absorption, and virtual corrections ensures the infrared finiteness of the decay rate at~$O(\alpha)$; all-order results for in-medium QED are given in~\cite{Weldon:1991eg,Indumathi:1996ec}. 

In the gravitational counterpart~\eqref{eq:vacuumprocess}--\eqref{eq:emissionprocess}, infrared sensitivity can be much greater due to the large graviton occupation numbers; the system is out of equilibrium, the radiation background is genuinely infrared, and perturbative cancellation alone may prove insufficient. 
Moreover, the QED case is simplified by the fact that screening by charged particles introduces an automatic infrared cutoff. Despite all matter being charged under gravity, the graviton does not acquire an in-medium mass.
To provide a definite answer and quantify gravitational corrections to candidate particle processes, we extend the soft-graviton %
theorem~\cite{Weinberg:1965nx} to include both real and virtual processes \emph{within} a gravitational background.
Resumming to all orders in~$G$, we demonstrate that inclusive decay rates remain—as expected—infrared safe, and that corrections to the vacuum process are minute.

This paper is organized as follows. We first construct the leading-order Boltzmann kinetic framework, then consider in turn real and virtual graviton corrections in the GW background, followed by their combination into an inclusive rate, and an interpretation of results. We use units of $\hbar = c = k_B= 1$.

\section{Soft graviton corrections I} 
The infrared (IR) safe rate for producing gravitons with total energy $\leq \bar E_{\rm em}$ in a process $\alpha \to \beta$ is given by~\cite{Weinberg:1965nx} %
\begin{align}
\label{eq:Msquared}
    d\Gamma_{\alpha \to \beta}^{(g)}(\leq \bar E_{\rm em})|_{\rm vac} = \left( \frac{\bar E_{\rm em}}{\Lambda} \right)^B b(B) d\Gamma_{\alpha \to \beta}^{0} \ .
\end{align}
Here, $d\Gamma_{\alpha\to \beta}^{0}$ is the differential rate for $\alpha \to \beta$ into some infinitesimal element of the momentum space of final state particles $\beta$ in the absence of gravitational corrections; $\Lambda$ is a fiducial scale that defines what is meant by ``soft gravitons'' and ensures the validity of~\eqref{eq:Msquared}, $\bar E_{\rm em}\leq \Lambda$ (see below).
The factor $B$ is given by
\begin{align}
\label{eq:Bfactor}
B=\frac{G}{2 \pi} \sum_{n m}  \frac{\eta_n \eta_m m_n m_m(1+\beta_{n m}^2)}{\beta_{n m}\left(1-\beta_{n m}^2\right)^{1 / 2}} \ln \left(\frac{1+\beta_{n m}}{1-\beta_{n m}}\right),
\end{align}
where $\beta_{nm}\equiv\left[1-m_n^2 m_m^2/(p_n\cdot p_m)^2\right]^{1/2}$ is the relative velocity of external leg particle $n$ and $m$ in the rest frame of either and $\eta_n=\pm 1$ for an outgoing (incoming)  particle of mass~$m_n$. The function $b(B) \simeq 1- \tfrac{1}{12}\pi^2 B^2$ can be approximated by unity for all processes we consider.
Taking the derivative yields $\frac{d}{d \bar E_{\rm em}} d\Gamma_{\alpha \to \beta}^{(g)}(\leq \bar E_{\rm em})|_{\bar E_{\rm em}=E;\, \rm vac} = (B/E)\e^{-B\ln(\Lambda/E)} d\Gamma_{\alpha \to \beta}^{0}$, which is the differential rate of emitting gravitons of total energy equal to $E$. Expanding it in powers of $B$ and taking the leading-order result, one can now identify the total energy $E$ as the energy of a single graviton $\omega$, and thus obtain the approximate rate for single graviton emission into an infinitesimal soft energy interval $[\omega, \omega + d\omega]$, $(B/\omega) d\Gamma_{\alpha \to \beta}^{0}$.
We have used the fact that $B\ln (\Lambda/\omega)\ll 1$ for essentially all frequency ranges of interest.  
In the Appendix, we show how equivalent rate expressions are obtained by explicit Feynman-diagrammatic calculations with the example of axion decay.

We now specialize to graviton corrections in the decay $\phi \to Y$ of a particle of mass $m_\phi$ in presence of a GW background. 
We may account for the additional possibilities of graviton absorption and stimulated emission through a Boltzmann-kinetic approach. The single-graviton emission rate above yields a general relation for the associated squared matrix elements
$ \int \mathrm{d}^2 \Omega_{\mathbf{q}}|\mathcal{M}_{\phi \rightarrow Y+g(\vecq)}|^2=\frac{4 \pi B}{\nu^2}|\mathcal{M}_{\phi \rightarrow Y}^0|^2$, 
where an integration over graviton direction~$\vec q$ has been effected and the graviton polarization sum is included. By crossing symmetry, this relation is also valid for single-graviton absorption with polarisation sum again included. These expressions are given in terms of the frequency $\nu = \omega/2\pi$ to facilitate the subsequent discussion of GW backgrounds.
The dimensionless $B$-factor ($B>0$ always) is, in general, a nontrivial function of the external momenta; for the important case of the two-body decay into photons, $\phi \to \gamma \gamma$, we find $B = (G/\pi)m_\phi^2$. More generally, one may note the useful parametric relation $B \sim (G/\pi)\mathcal Q^2$, where $\mathcal Q$ denotes the mass-energy ($Q$-value) released in the decay of~$\phi$.
Using the squared matrix elements in a standard Boltzmann equation for $\phi$-decay in a non-degenerate, dilute medium---so that the only relevant occupation numbers $f_i$ are those of $\phi$ and $g$---one obtains for the change of the $\phi$ number density $n_\phi$ (neglecting inverse decays and momentum dependence of~$B$)
\begin{align}
\label{eq:enhancedDecay}
     \frac{dn_{\phi}}{dt} = - n_\phi \Gamma_{\phi\rightarrow Y}^0\left[ 1+ 2 B  \int_{\nu_{\rm min}}^{\Lambda} d\ln\nu\ f_g(\nu) \right].
\end{align}
Here, $f_g$ is the graviton occupation number and $\nu_{\rm min}$ is an infrared frequency cutoff set either by the GW spectrum or by the collision time of $\phi$ or~$Y$ in the medium.
The factor of~$2$ arises because the rates for graviton absorption and stimulated emission are equal.
Since $f_g \gg 1$, we have neglected spontaneous graviton emission in~\eqref{eq:enhancedDecay}.

Typical sources for GWs, whether stochastic or transient, generate an enormous number of quanta in the low-frequency regime. Let $\rho_g$ denote the GW energy density. We have $\d \rho_g/\d \ln \nu=\pi h_c^2 \nu^2/(4G)$ where $h_c$ is the characteristic strain. On the other hand, $\rho_g$ can be related to the graviton occupation number as $\rho_g=2\int d^3\vecq\, |\vecq| f_g(|\vecq|)/(2\pi)^3$ with $|\vecq|=2\pi\nu$. We hence obtain $f_g=h_c^2/(64\pi G\nu^2)$.
Adopting values representative of the amplitude and frequency suggested by pulsar timing array observations, the enhancement factor in~\eqref{eq:enhancedDecay} evaluates to an implausibly large value,
\begin{align}
\label{eq:Bfvalue}
    2B \int_{\nu_{\rm min}}^{\Lambda} d\ln\nu\ f_g \simeq %
    10^{14}
    \left( \frac{\mathcal Q}{\rm eV}\right)^2 \left( \frac{h_c}{10^{-15}}\right)^2 \left( \frac{\text{nHz}}{\nu_{\rm min}}\right)^2 ,
\end{align} 
where, for simplicity, we have assumed a frequency-independent characteristic strain.
Rather than indicating a genuine enhancement of the decay rate, this instead signals that we have left the regime of validity of a leading-order treatment in~$G$. In the remainder of this work, we demonstrate that~\eqref{eq:enhancedDecay} is incorrect, and that a proper treatment requires revisiting the soft-graviton theorem in the presence of a GW background. This constitutes our central result.

Before proceeding, we comment on~\cite{Flauger:2019cam}. 
In that work, the graviton emission and absorption rates in a medium of temperature $T$ are considered, notably from the non-relativistic elastic scattering of electrons with protons.%
\footnote{We note in passing a missing term $(-11/6)\vec v_m^2 \vec v_n^2$ in (4.10) of~\cite{Weinberg:1965nx} which yields instead $B=(G / \pi) (16 / 5)  \left[Q_{i j} Q_{i j}- (1/3) \left(Q_{i i}\right)^2\right]$ for~(4.13); see~\cite{Garcia-Cely:2024ujr} for a further discussion on the associated rates.} 
Each of the energy-differential rates comes with a factor $B f_g$ that evaluates to the right hand side of~\eqref{eq:Bfvalue} when substituting $\nu_{\rm min}\to \nu$. Owing to the fact that $\mathcal{Q} \sim T \gg \text{eV}$ for the elastic processes they consider, even larger values are obtained.
Crucially for their result, as they examine the evolution of GWs rather than that of the interacting fields, the absorption and emission processes carry \emph{opposite} signs. 
As a result, stimulated processes cancel almost exactly, up to a factor of $e^{-\omega/T}$ that arises from the thermal equilibrium of the medium. The partial cancellation leads to the correct qualitative conclusion; as we shall see, our all-orders treatment corroborates this cancelation and places the transparency of matter to GWs on firmer footing.
In this work, we identify a counterexample: particle decays furnish a scenario in which both stimulated rates contribute with the \emph{same} sign when the abundance of~$\phi$ is concerned, and no such cancellation occurs. 
Consequently, going beyond the current treatment and incorporating the GW background into~\eqref{eq:Msquared} is required to obtain consistent results.

\section{In-medium virtual graviton corrections} We first introduce the differential vacuum transition rate $d{\Gamma}_{\alpha\rightarrow\beta}^0(p_1,...,p_{N_\alpha+N_\beta})$ for a process $\alpha (p_{1},...p_{N_\alpha})\rightarrow\beta(p_{N_\alpha+1},...p_{N_\alpha+N_\beta}) $ in the absence of gravitational interactions.
Attaching a soft graviton of four-momentum~$q$ to an external leg~$n$, amounts to multiplying the eikonal factor $S^{\mu\nu}_n = \frac{\sqrt{32\pi G}\,\eta_n p_n^\mu p^\nu_n}{2p_n\cdot q+\ii\eta_n \epsilon}$~\cite{Weinberg:1965nx}. 
Connecting two particles $m,n$ through a virtual graviton, amounts to supplying two soft factors $S_m^{\mu\nu}$ and $S_n^{\rho\sigma}$ together with the graviton propagator and performing the integral over~$d^4q$. In the presence of a homogeneous graviton background, the \emph{in-medium} propagator becomes, in the de Donder gauge, $\Delta_{\mu\nu,\rho\sigma}(q) = P_{\mu\nu\rho\sigma}\big( \frac{\ii }{q^2+\ii\epsilon} + 2\pi \delta(q^2)\left[\theta(q^0)f_g(\vecq)+\theta(-q^0)f_g(-\vecq)\right] \big)$ where the second term represents the on-shell, statistical contribution of the gravitational background;
$P_{\mu\nu\rho\sigma}=\frac{1}{2}\left(\eta_{\mu\rho}\eta_{\nu\sigma}+\eta_{\mu\sigma}\eta_{\nu\rho}-\eta_{\mu\nu}\eta_{\rho\sigma}\right)$. 
A virtual soft-graviton exchange is therefore given by the factor $\int_\lambda^\Lambda \frac{\d^4 q}{(2\pi)^4} S^{\mu\nu}_m \Delta_{\mu\nu,\rho\sigma}(q) S^{\rho\sigma}_n$,
where we have introduced~$\Lambda$ as a dividing point justifying the eikonal approximation and defining a soft graviton. The  $\Lambda$-dependence in the rates is cancelled by including hard virtual graviton corrections into~$d{\Gamma}_{\alpha\rightarrow\beta}^0$; since $ B\ln(m_\phi/\Lambda)\ll 1$, neglecting the latter remains a good approximation, see~\cite{Weinberg:1995mt}.
The IR cutoff $\lambda$ on $\omega\equiv|\vecq|$ will cancel later, once real processes are taken into account. Attaching $N$ virtual soft gravitons in the process $\alpha\to \beta$ leads to a factor of
\begin{align}
    &\frac{1}{2^N N!}\prod_{i=1}^N \left(\sum_{m,n}\int_\lambda^\Lambda \frac{\d^4 q_i}{(2\pi)^4} S^{\mu_i \nu_i}_{m} \Delta_{\mu_i \nu_i,\rho_i\sigma_i}(q_i) S^{\rho_i\sigma_i}_{n}\right)\notag\\
    &=\frac{1}{N!}\left(\frac{1}{2}\int_\lambda^\Lambda \frac{\d^4 q}{(2\pi)^4}\widetilde{B}(q)\right)^N\,,
\end{align}
where 
\begin{align*}
    &\widetilde{B}(q)=
    8\pi G\left(\sum_{m,n}\frac{\eta_m\eta_n \left[(p_m\cdot p_n)^2-\frac{1}{2}m_m^2 m_n^2\right]}{(p_n\cdot q+\ii\eta_n \epsilon)(-p_m\cdot q+\ii\eta_m\epsilon)}\right)\notag\\
    &\left(\frac{\ii}{q^2+\ii\epsilon}+2\pi\delta(q^2) \left[\theta(q^0)f_g(\vecq,x)+\theta(-q^0)f_g(-\vecq,x)\right] \right) .
\end{align*}
Summing over $N$ from zero to infinity, translates into a rate
$
    d{\Gamma}_{\alpha\rightarrow\beta}=\exp\left[{\rm Re}\int_\lambda^\Lambda \frac{\d^4 q}{(2\pi)^4} \widetilde{B}(q) \right] d{\Gamma}_{\alpha\rightarrow\beta}^0 .
$ 
Both, the vacuum and in-medium parts of $\widetilde{B}(q)$ contribute to the real part in the exponential of the integral and we find, 
\begin{align*}
    {\rm Re} \int_\lambda^\Lambda \frac{\d^4 q}{(2\pi)^4} \widetilde{B}(q) =-B\int_\lambda^\Lambda\frac{\d\omega}{\omega} \left[1+2f_g(\omega)\right]\,, 
\end{align*}
where $B=\int\d^2\Omega_{\hat q}\, \oldhat{B}(\hat{q}) $ is the integral over the directions of $\hat q = \vec q/|\vec q|$ with
$\oldhat{B}(\hat{q})$ being a standard expression found as Eq.~(2.25) in~\cite{Weinberg:1965nx}.
In the above derivation, we have assumed that the graviton background is isotropic, i.e., $f_g(\vecq)=f_g(|\vecq|)\equiv f_g(\omega)$. 
Integrating over the solid angle, one recovers~\eqref{eq:Bfactor}.
In conclusion, we now have 
\begin{align}
\label{eq:Gamma}
    d\Gamma_{\alpha\rightarrow\beta}
    =\,&d\Gamma^0_{\alpha\rightarrow\beta}%
    \left(\frac{\lambda}{\Lambda}\right)^B \exp\left[-2B\int_\lambda^\Lambda\frac{\d\omega}{\omega} f_g(\omega)\right]\,.
\end{align}
The last factor is the modification due to the presence of the gravitational background. Note that virtual corrections suppress the rate~$d\Gamma^0$.

\section{In-medium graviton absorption and emission} Suppose we consider the rate for 
producing $N_{\rm em}$ and absorbing $N_{\rm abs}$ soft gravitons with respective momenta near $\vec q_1,\dots,\vec q_{N_{\rm em}}$ and $\vec q_{N_{\rm em}+1},\dots,\vec q_{N_{\rm em}+N_{\rm abs}}$.
To the $S$-matrix element, the $i$-the emitted graviton will contribute a factor of $\sqrt{8\pi G}(2\pi)^{-3/2}(2|\vecq_i|)^{-1/2}[1+f_g(\vecq_i)]^{1/2} \sum_n\eta_n[p_n\cdot \epsilon^*(\vecq_i,\frac{1}{2}h_i )]^2/(p_n\cdot q_i)$, and the $j$-th absorbed graviton will contribute a factor of $\sqrt{8\pi G}(2\pi)^{-3/2}(2|\vecq_j|)^{-1/2}f_g(\vecq_j)^{1/2} \sum\eta_n[p_n\cdot \epsilon(\vecq_j,\frac{1}{2}h_j )]^2/(-p_n\cdot q_j)$. Here $h_{i,j}=\pm 2$, and we have used the fact that the graviton polarisation tensor can be written as $\epsilon_{\mu\nu}(\vecq,\pm)=\epsilon_\mu(\vecq,\pm)\epsilon_\nu(\vecq,\pm)$. Compared with the form given in~\cite{Weinberg:1965nx}, we now have inserted factors of the graviton occupation number.  Taking the absolute square of the $S$-matrix element, summing over helicities and dividing by $N_{\rm em}!$ for emission and $N_{\rm abs}!$ for absorption, we obtain the desired rate,
\begin{widetext}
\begin{align}
\label{eq:Gamma-general}
     d{\Gamma}_{\alpha\rightarrow\beta}^{(g)}(q_1,...q_{N_{\rm em}+N_{\rm abs}})%
     =d{\Gamma}_{\alpha\rightarrow\beta} \left(\frac{1}{N_{\rm em}!}\prod_{i=1}^{N_{\rm em}} \mathscr{B}(\vecq_i) [1+f_g(\vecq_i)]\d^3\vecq_i\right)\left(\frac{1}{N_{\rm abs}!}\prod_{j=N_{\rm em}+1}^{N_{\rm em}+N_{\rm abs}} \mathscr{B}(\vecq_j) f_g(\vecq_j)\d^3\vecq_j\right) ,
\end{align}
\end{widetext}
where $\mathscr{B}(\vecq)$ is given by~\cite{Weinberg:1965nx}
$
    \mathscr{B}(\vecq)= \frac{8\pi G}{(2\pi)^3 2|\vecq|}
    \sum_{m,n} \frac{\eta_m\eta_n p_m^\mu p_m^\nu p_n^\rho p_n^\sigma \Pi_{\mu\nu\rho\sigma}(q)}{(p_m\cdot q)(p_n\cdot q)}
$
with $\Pi_{\mu\nu\rho\sigma}(q)=\sum_{\pm } \epsilon_{\mu}(\vecq,\pm)\epsilon_\nu(\vecq,\pm) \epsilon^*_\rho(\vecq,\pm) \epsilon^*_\sigma(\vecq,\pm)$.
Above, we have suppressed the hard particle momentum labels as before.  On-shell conditions are understood for the four-momenta $q_{i,j}$, and $\omega_{i,j}\equiv q^0_{i,j}\geq 0$.
Note that if either $N_{\rm em}$ or $N_{\rm abs}$ are zero, the corresponding factor multiplying $d\Gamma_{\alpha\to\beta}$ in Eq.~\eqref{eq:Gamma-general} should be set to unity.
The same logic should be applied to Eq.~\eqref{eq:dGammaOmegas} below.

We recall that\,\cite{Weinberg:1965rz,vanDam:1970vg} $\Pi_{\mu\nu\rho\sigma}(q)=\frac{1}{2}[\hat{\eta}_{\mu\rho}(q)\hat{\eta}_{\nu\sigma}(q)+\hat{\eta}_{\mu\sigma}(q)\hat{\eta}_{\nu\rho}(q)-\hat{\eta}_{\mu\nu}(q)\hat{\eta}_{\rho\sigma}(q)]
$ with $\hat{\eta}_{\mu\nu}=\eta_{\mu\nu}-\frac{q_{\mu} \bar{q}_{\nu}+\bar{q}_{\mu} q_{\nu}}{q\cdot\bar{q} },\ \bar{q}=(q^0,-\vecq)$.
Due to four-momentum conservation $\sum\eta_n p_n^\mu=0$ in the soft limit, the terms with $\bar{q}$ do not contribute and we can take $\Pi_{\mu\nu\rho\sigma}\rightarrow P_{\mu\nu\rho\sigma}$. Therefore, we have $\mathscr{B}(\vecq)=\oldhat{B}(\hat{q})/|\vecq|^3$.

Again, assuming that the graviton background is isotropic, we can integrate over the solid angles in~\eqref{eq:Gamma-general} and obtain 
\begin{widetext}
\begin{align}
\label{eq:dGammaOmegas}
    d\Gamma_{\alpha\rightarrow\beta}^{(g)}(\omega_1,...,\omega_{N_{\rm em}+N_{\rm abs}})
    =d \Gamma_{\alpha\rightarrow\beta} \left(\frac{B^{N_{\rm em}}}{N_{\rm em}!}\prod_{i=1}^{N_{\rm em}}\frac{\d \omega_i}{\omega_i} [1+f_g(\omega_i)] \right)\left(\frac{B^{N_{\rm abs}}}{N_{\rm abs}!}\prod_{j=N_{\rm em}+1}^{N_{\rm em}+N_{\rm abs}}\frac{\d \omega_j}{\omega_j} f_g(\omega_j) \right)\,.
\end{align}
\end{widetext}

Following~\cite{Weinberg:1965nx}, we now consider inclusive rates with an arbitrary number of emitted gravitons of total energy no more than $\bar E_{\rm em}$, and with an arbitrary number of absorbed gravitons of total energy no more than $\bar E_{\rm abs}$. We have
$
    d\Gamma_{\alpha\rightarrow\beta}^{(g)}(\leq \bar E_{\rm em},\leq \bar E_{\rm abs})=\sum_{N_{\rm em},N_{\rm abs}=0}^{\infty}\int_\lambda^{\bar E_{\rm em}}%
    \int_\lambda^{\bar E_{\rm abs}}%
    d\Gamma_{\alpha\rightarrow\beta}^{(g)}(\omega_1,...,\omega_{N_{\rm em}+N_{\rm abs}})
    \Theta_{\rm em} \Theta_{\rm abs}\,,
$
where the energy-restricting theta-functions are given by $\Theta_{\rm em} =  \Theta\left(\bar E_{\rm em}-\sum_{i=1}^{N_{\rm em}}\omega_i\right)$ for emission and $\Theta_{\rm abs} = \Theta\left(\bar E_{\rm abs}-\sum_{j=N_{\rm em}+1}^{N_{\rm em}+N_{\rm abs}}\omega_j\right)$ for absorption.
Substituting the representation of the step function, 
$%
\Theta_{\rm em}
=\frac{1}{\pi} \int_{-\infty}^\infty \d\sigma\frac{\sin(\bar E_{\rm em}\sigma)}{\sigma} \, \e^{\ii\sigma\sum_{i=1}^{N_{\rm em}}\omega_i}$ and similarly for $\Theta_{\rm abs}$,
we then obtain 
\begin{align}
\label{eq:Gamma-g2}
    &d\Gamma^{(g)}_{\alpha\rightarrow\beta}(\leq \bar E_{\rm em}, \leq\bar E_{\rm abs}) =
 \e^{B\left(\int_{\lambda}^{\bar E_{\rm em}}\frac{\d\omega}{\omega} f_g(\omega)+\int_{\lambda}^{\bar E_{\rm abs}}\frac{\d\omega}{\omega} f_g(\omega)\right)}
 \notag\\ &
    \left(\frac{\bar E_{\rm em}}{\lambda}\right)^B b_{\rm em}(B,\bar E_{\rm em};f_g)b_{\rm abs}(B,\bar E_{\rm abs};f_g) 
    d\Gamma_{\alpha\rightarrow\beta} ,
\end{align}
where
  \begin{align*}
    &b_{\rm em}(x,\bar E_{\rm em};f_g)=\int_{-\infty}^\infty \d\sigma\frac{\sin\sigma}{\pi\sigma}\, \e^{x\int_0^1 \frac{\d y}{y}\left[1+f_g\left(y \bar E_{\rm em}\right)\right]\left(\e^{\ii y\sigma}-1\right)}\,,
\end{align*}  
and a similar expression for $b_{\rm abs}(x,\bar E;f_g)$, obtained from $b_{\rm em}(x,\bar E;f_g)$ by replacing $\left[1+f_g\left(y \bar E\right)\right]$ by $f_g\left(y \bar E\right)$ in the exponential.
Note that for $f_g=0$, $b_{\rm abs}(x,\bar E_{\rm abs};f_g=0)=1$ and $b_{\rm em}(x,\bar E_{\rm em};f_g=0)$ reduces to the function $b(x)$ given in~\cite{Weinberg:1965nx}.

\section{Final inclusive rates and interpretation} Combining~\eqref{eq:Gamma} and~\eqref{eq:Gamma-g2}, we finally obtain
the rate for a hard process $\alpha \to \beta$ to emit and absorb gravitons with energies no greater than $\bar E_{\rm em}$ and $\bar E_{\rm abs}$, respectively:
\begin{align}
\label{eq:new-soft-theorem}
    &d\Gamma^{(g)}_{\alpha\rightarrow\beta}(
    \leq \bar E_{\rm em}, \leq \bar E_{\rm abs}) =d\Gamma^0_{\alpha\rightarrow\beta}\,\times\notag\\
    &
    b_{\rm em}(B, \bar E_{\rm em};f_g) b_{\rm abs}(B, \bar E_{\rm abs};f_g)\mathcal{R}(\bar E_{\rm em}, \bar E_{\rm abs};f_g)
     \,,
\end{align}
where
\begin{align*}
    \mathcal{R}(\bar E_{\rm em}, \bar E_{\rm abs};f_g)=\e^{-B\left(\int^{\Lambda}_{\bar E_{\rm em}}\frac{\d\omega}{\omega} [1+f_g(\omega)]+\int^{\Lambda}_{\bar E_{\rm abs}}\frac{\d\omega}{\omega} f_g(\omega)\right)} \ .
\end{align*}
This is our central result. It extends the soft-graviton theorem in the presence of a gravitational background. 
For $f_g=0$, we recover Weinberg's original expression~\cite{Weinberg:1965nx}. Note that the IR-cutoff $\lambda$ cancels in the above expression. Equation~\eqref{eq:new-soft-theorem} deserves some comments. 

First, in typical scenarios, the factors $b_{\rm em/abs} \simeq 1$. The factor $ \mathcal{R}(\bar E_{\rm em}, \bar E_{\rm abs};f_g) \in [0,1]$ can then be interpreted as the probability that 
the total emitted and absorbed graviton energies are less than $\bar E_{\rm em}$ and $\bar E_{\rm abs}$, respectively.
Taking $\bar E_{\rm em} \to 0$ causes $d\Gamma^{(g)}_{\alpha\rightarrow\beta}$ to vanish, recovering the well-known result that the probability of emitting no graviton energy is infinitely suppressed. For $f_g=0$, this reproduces the well-known
statement that in any hard process soft gravitons are inevitably emitted~\cite{Bloch:1937pw,Yennie:1961ad,Kinoshita:1962ur,Lee:1964is}.
Meanwhile, \emph{absorption} leads to an additional suppression of $d\Gamma^{(g)}_{\alpha\rightarrow\beta}$ due to the presence of a gravitational wave background that is most pronounced when $\bar E_{\rm abs}\to 0$ .
Operationally, one may regard $\bar E_{\rm em} $ and $\bar E_{\rm abs}$ as the soft-energy thresholds of a detector; in that sense, $\bar E_{\rm em}+\bar E_{\rm abs}$ plays the role of an effective energy resolution.

In cosmological applications, e.g.~DM decay such as $\phi\to \gamma\gamma$, the same logic applies. The detector's energy resolution is now the width of a frequency bin: $\bar E_{\rm em}+\bar E_{\rm abs}$ represents the total frequency interval within which photons are counted as indistinguishable. Neglecting cosmological redshift and other factors leading to a broadening of the line shape, the $\phi$-decay at rest produces a line at $m_\phi/2$; any photon within $[m_\phi/2-\bar E_{\rm em}, m_\phi/2+ \bar E_{\rm abs}]$ then contributes to the same observed line.
If the observable is not directly tied to the final states of the scattering process---e.g.~if we are only interested in the total $\phi$ abundance---all processes with any allowed emitted or absorbed graviton energy must be included (fully inclusive rate). This amounts to setting $\bar E_{\rm em}=\bar E_{\rm abs}=\Lambda$, yielding $d\Gamma^{(g)}_{\alpha\rightarrow\beta} =d\Gamma^0_{\alpha\rightarrow\beta}b_{\rm em}(B,\Lambda;f_g)b_{\rm abs}(B,\Lambda;f_g)\approx d\Gamma^0_{\alpha\rightarrow\beta}$.
We now explicitly see that~\eqref{eq:enhancedDecay} is misguided: the naive vacuum decay rate of~$\phi$ remains essentially unaltered.

Second, we are now in a position to return to the general question of Bose-enhanced emission rates in a GW background. We may again try to anchor this discussion using the IR-safe energy-restricted rate Eq.~\eqref{eq:new-soft-theorem}, 
and differentiate it with respect to its threshold. In vacuum, the rate is~\eqref{eq:Msquared}, and its derivative after dropping $\e^{-B\ln(\Lambda/E)}\simeq 1$ is  
the single-graviton spectrum giving~\eqref{eq:enhancedDecay} and used in~\cite{Flauger:2019cam}.
Taking the gravitational background into account, the IR-safe rate is now~\eqref{eq:new-soft-theorem}.  Differentiating it, we now must keep the exponential as it can be far from unity.
Marginalizing over absorbed gravitons 
by setting $\bar E_{\rm abs} = \Lambda$  yields
\begin{align}
\label{eq:emission-rates}
    \bar E_{\rm em}\frac{d}{\d \bar E_{\rm em}}\d \Gamma^{(g)}_{\alpha\rightarrow\beta}(\leq \bar E_{\rm em},\leq \Lambda)|_{\bar E_{\rm em}=E }= E S'(E) d\Gamma^0_{\alpha\rightarrow\beta} \ ,
\end{align}
where we neglected the unity factors $b_{\rm em}$ and $b_{\rm abs}$.
The function $
    S(E) \equiv \mathcal{R}(E, \Lambda;f_g) = \e^{-B\int^{\Lambda}_{E}\frac{\d\omega}{\omega}[1+f_g(\omega)] }$
acts as an in-medium Sudakov factor, and its derivative
\begin{align}
\label{eq:exclusive-rate}
 S'(E) = \frac{B}{E}[1+f_g(E)]  S(E) \ ,
\end{align}
is the probability distribution for emitting  gravitons of \emph{combined} total energy $E$ in the process $\alpha \to \beta$. By construction, $\int_0^\Lambda dE\, S'(E) = 1$.
A Bose-enhanced rate (in a logarithmic energy interval) requires $E S'(E) \gg 1$. For $B\int_{E}^\Lambda \frac{d\omega}{\omega} [1+f_g(\omega)]\ll 1$, one finds $E S'(E)\approx B[1+f_g(E)]$, which corresponds to the conventional form of Bose enhancement.  However, when $B\int_{E}^\Lambda \frac{d\omega}{\omega} [1+f_g(\omega)]\gg 1$, $E S'(E)$ becomes strongly suppressed: 
it is exponentially rare that the \emph{total} emitted energy in the hard process is as small as~$E$. In the situation \eqref{eq:Bfvalue}, the background stimulates emission 
so that a hard process $\alpha\to\beta$ radiates vastly more; the suppression of $S'$ is a statement about the summed energy of all
emitted quanta, not about emission at frequency~$E$.

Indeed, physical emission and absorption spectra are {\it not} suppressed. Equation~\eqref{eq:dGammaOmegas} describes two Poisson processes with intensities $\mu_{\rm em}(\omega)=(B/\omega)[1+f_g(\omega)]$ and $\mu_{\rm abs}(\omega)=(B/\omega)f_g(\omega)$, endowed by the correct normalization through the virtual corrections of Eq.~\eqref{eq:Gamma}. Summing over unobserved companion quanta, the mean number of gravitons emitted and absorbed into $[\omega, \omega +d\omega]$ in a hard process $\alpha\to \beta$ is (see Appendix)
\begin{align}
\label{eq:expected-numbers}
    \frac{d\langle N_{\rm em}\rangle}{d\omega}=\frac{B}{\omega}\left[1+f_g(\omega)\right],
    \quad
    \frac{d\langle N_{\rm abs}\rangle}{d\omega}=\frac{B}{\omega}f_g(\omega)\,,
\end{align}
independent of $\lambda$ and $\Lambda$, and to all orders in $G$. We conclude that Bose-enhancement of the soft graviton absorption and emission survives resummation and retains its leading-order functional form. What the resummation removes is the imprint on the total rate for $\alpha\to \beta$: the would-be enhancement in~\eqref{eq:enhancedDecay} is cancelled by the in-medium virtual corrections. Since $S'(\omega)=[\d\langle N_{\rm em}\rangle/\d\omega]\,S(\omega)$, only in vacuum
does~\eqref{eq:emission-rates} coincide with the emission spectrum as there one has  $S\simeq 1$
for any conceivable hierarchy of scales.

This demonstrates the necessity of the in-medium soft-graviton theorem~\eqref{eq:new-soft-theorem}: only by consistently accounting for the medium-modified infrared structure 
do real and virtual exchanges combine into well-defined observables. For the propagation of GW themselves, the relevant quantity is the difference of the mean spectra~\eqref{eq:expected-numbers}, gaining ${d\langle N_{\rm em}\rangle}/{d\omega}$ and losing ${d\langle N_{\rm abs}\rangle}/{d\omega}$ quanta at frequency $\omega$ per hard process, so that the Bose-enhanced parts cancel \emph{identically}. In a thermal medium at temperature $T$, this exact cancellation is lifted once the asymmetry of hard rates by a factor $\e^{\omega/T}$ is taken into account. The cancellation identified in~\cite{Flauger:2019cam} at lowest order therefore persists to all orders in~$G$ and at arbitrary graviton multiplicity. The residual attenuation is controlled by the detailed balance of the medium, which softens $\Gamma_{\rm abs}\propto\nu^{-3}$ to $\Gamma_{\rm abs}(1-\e^{-\omega/T})$~\cite{Flauger:2019cam}; a complete radiative-transfer analysis, executed to all orders in~$G$, will be presented elsewhere~\cite{upcoming}.

Formally, the in-medium factor  $\mathcal{R}$ in~\eqref{eq:new-soft-theorem}  has the same structure as known from thermal QED~\cite{Weldon:1991eg}: one may obtain it by the replacements $B\to A$ with $A$ defined by~(2.16) in~\cite{Weinberg:1965nx} and $f_g\to f_\gamma$ where $f_\gamma$ is the photon occupation number. 
Equation~\eqref{eq:new-soft-theorem} is the gravitational completion of the finite-temperature Bloch–Nordsieck/Yennie-Frautschi-Suura~resummation~\cite{Bloch:1937pw,Yennie:1961ad}, providing the first explicit in-medium soft-graviton theorem applicable for any state of soft gravitons, including thermal and coherent states with possibly large occupancies.
In contrast with the QED case, the need for the in-medium soft-graviton theorem is more acute due to the lack of screening experienced by gravitons.
Our theorem therefore renders graviton processes IR safe even in the regime where the graviton energy is the smallest scale in the problem.

\section{Conclusions} 
The combination of ultra-soft frequencies and enormous graviton occupation numbers forces a reassessment of what is meant by ``weak coupling'' in gravity. As we have seen, at the level of the Boltzmann equation, one may easily be led to expect large Bose factors to enhance decay, emission, or absorption rates, with putative implications for nHz GW transparency, DM stability, or other cosmological processes. Yet, as follows from Eq.~\eqref{eq:new-soft-theorem}, once real and virtual exchanges are treated on equal footing and the gravitational background itself is taken into account,
the fate of these Bose factors depends on the observable: in inclusive rates~\eqref{eq:new-soft-theorem} they cancel, in the mean emission and absorption spectra~\eqref{eq:expected-numbers} they persist, and in inclusive rates with specified total emitted soft-graviton energy~\eqref{eq:emission-rates} they exponentiate into a suppression. The in-medium Sudakov factor $\mathcal{R}(\bar E_{\rm em},\bar E_{\rm abs};f_g)$ contains the full infrared bookkeeping; its exponential is the normalization of the multi-graviton emission and absorption probabilities.

Weinberg's ``Second Law of Progress in Theoretical Physics'' cautions  us against trusting arguments based on the lowest order of perturbation theory~\cite{Weinberg:1983RG}. The supposed Bose enhancement of the total decay rate
is such a lowest-order mirage. Only after resumming the complete series of soft emissions and absorptions in a gravitational background does the physically consistent picture emerge: 
every decay exchanges an enormous number of soft gravitons with the background, yet proceeds at its vacuum rate. The Bose enhancement resides in the emission and absorption spectra alone, where it cancels between the two channels, leaving long-wavelength GW essentially unimpeded.
That the weakest of all forces demands such treatment is, perhaps, its own quiet vindication of that law.

\begin{acknowledgments}
We thank Diego Blas and Francesco Riva for useful discussions
and Ryan Plestid and an anonymous referee for critical comments
that improved the manuscript.
Funded/Co-funded by the European Union (ERC, NLO-DM, 101044443) and by the SNF (Ambizione grant PZ00P2\_193322). This work was also supported by the Research Network Quantum Aspects of Spacetime (TURIS) and in part by grant NSF PHY-2309135 to the Kavli Institute for Theoretical Physics (KITP).
\end{acknowledgments}

\appendix

\section{Axion decay in association with graviton absorption and emission}

In this appendix, we show how a leading-order treatment in the gravitational corrections to $\phi$ decay yields a seemingly enhanced decay rate in the presence of a gravitational background. We illustrate this on the example of axion decay to two photons, $\phi\to \gamma \gamma$.
To this end, we first compute the squared transition amplitude using an explicit Feynman-diagrammatic calculation and subsequently take the soft-graviton limit to obtain the enhancement factors quoted in the main text. As a consistency check, we then verify that the same result follows directly from Weinberg's original soft-graviton theorem. The fact that the results obtained this way are not correct demonstrates the necessity of generalising Weinberg’s soft-graviton theorem to scenarios involving a background medium of soft gravitons. 

\subsection{Leading-order Feynman-diagrammatic calculation} 
We start from the Lagrangian describing an axion-like field~$\phi$ coupled to electromagnetism in a dynamical spacetime background,
\begin{align}
    \L= &\sqrt{-g}\left[\frac{1}{2}(\partial_\mu \phi) \partial^\mu\phi -\frac{1}{2}m_\phi^2\phi^2 -\frac{1}{4} F_{\alpha\beta} F^{\alpha\beta}\right] \notag \\ & -\frac{g_{\phi\gamma}}{4} \phi F_{\mu\nu} \widetilde{F}^{\mu\nu}\,.
\end{align}
where $F_{\mu\nu}$ is the photon field strength tensor and
$\widetilde F^{\mu\nu} = \tfrac{1}{2} \epsilon^{\mu\nu\rho\sigma} F_{\rho\sigma}$ is its dual. Note that the last term is topological and does not induce a direct interaction involving gravitons. After the expansion $g_{\mu\nu}=\eta_{\mu\nu}+\kappa h_{\mu\nu}$ where $\eta_{\mu\nu}$ is the flat Minkowski metric,  to  leading order in $\kappa=2/M_{\rm pl}=2\sqrt{8\pi G}$, this Lagrangian yields the scalar-graviton, photon-graviton and photon-scalar three-vertices:%
\tikzset{
  every picture/.style={scale=0.8, every node/.style={transform shape}},
}
\tikzfeynmanset{
  momentum/arrow distance=2.5mm,
  momentum/arrow shorten=0.20,
}
\newcommand{\diagraise}{-0.4\height} 
\begin{align*}
\raisebox{\diagraise}{%
\begin{tikzpicture}[baseline={(current bounding box.center)}]
\begin{feynman}
    \vertex (a) at (-1.4,0) {\(\phi\)};
    \vertex (b) at (0,0);
    \vertex (c) at ( 1.4,0) {\(\phi\)};
    \vertex (f) at (0,1.5) {\(h_{\mu\nu}\)};
    \diagram*{
      (a) -- [scalar,  momentum'={\(p_1\)}] (b)
          -- [scalar,  momentum'={\(p_2\)}] (c),
      (b) -- [graviton, shorten <=2pt, shorten >=2pt] (f),
    };
  \end{feynman}
\end{tikzpicture}}
\raisebox{\diagraise}{%
\begin{tikzpicture}[baseline={(current bounding box.center)}]
  \begin{feynman}
    \vertex (a) at (-1.4,0) {\(A_{\alpha}\)};
    \vertex (b) at (0,0);
    \vertex (c) at ( 1.4,0) {\(A_{\beta}\)};
    \vertex (f) at (0,1.5) {\(h_{\mu\nu}\)};
    \diagram*{
      (a) -- [photon,  momentum'={\(p_1\)}] (b)
          -- [photon,  momentum'={\(p_2\)}] (c),
      (b) -- [graviton, shorten <=2pt, shorten >=2pt] (f),
    };
  \end{feynman}
\end{tikzpicture}}
\raisebox{\diagraise}{%
\begin{tikzpicture}[baseline={(current bounding box.center)}]
  \begin{feynman}
    \vertex (a) at (-1.4,0) {\(A_{\alpha}\)};
    \vertex (b) at (0,0);
    \vertex (c) at ( 1.4,0) {\(A_{\beta}\)};
    \vertex (f) at (0,1.5)  {\(\phi\)};
    \diagram*{
      (a) -- [photon, momentum'={\(p_1\)}] (b)
          -- [photon, momentum'={\(p_2\)}] (c),
      (b) -- [scalar, shorten <=2pt, shorten >=2pt] (f),
    };
  \end{feynman}
\end{tikzpicture}}\,.
\end{align*}
The corresponding Feynman rules read~\cite{Choi:1994ax}
\begin{subequations}
\begin{align}
&\text{$\phi\phi h$:}\quad -\frac{i\kappa}{2} \Big[p_1^\mu p_2^\nu +p_1^\nu p_2^\mu -\eta^{\mu\nu} (p_1\cdot p_2-m^2_\phi)\Big]\ , \\
&\text{$\gamma\gamma h$:}\quad  -\frac{i\kappa}{2}\Big[ \eta^{\nu\beta} p_1^\mu p_2^\alpha
-\eta^{\alpha\beta} p_1^\mu p_2^\nu  
+\eta^{\mu\alpha}p_1^\beta p_2^\nu\notag\\
&\quad\qquad-\eta^{\mu\nu}p_1^\beta p_2^\alpha +\eta^{\nu\alpha}p_1^\beta p_2^\mu
-\eta^{\alpha\beta}p_1^\nu p_2^\mu +\eta^{\mu\beta} p_1^\nu p_2^\alpha\notag\\   
&\quad\qquad+\eta^{\mu\nu}\eta^{\alpha\beta} p_1\cdot p_2 -\eta^{\nu\alpha}\eta^{\mu\beta} p_1\cdot p_2 \notag
\\ &
\quad\qquad-\eta^{\nu\beta}\eta^{\mu\alpha} p_1\cdot p_2 \Big]\,,\\
&\text{$\gamma\gamma\phi$:} \quad i g_{\phi\gamma}\, \epsilon^{\alpha\beta\mu\nu}\, p_{1\mu}\, p_{2\nu}\,.
\end{align}    
\end{subequations}

\begin{figure*}[t]
\centering
\begin{tikzpicture}[scale=0.9,baseline={-0.025cm*height("$=$")}]
  \begin{feynman}
    \vertex (a) at (-2.5,-1.5) {\(\phi\)};
    \vertex (b) at (-1,0) ;
    \vertex (c) at (-2.5,1.5) {\(h_{\mu\nu}\)} ;
    \vertex (d) at (1,0);
    \vertex (e) at (2.5,-1.5) {\(A_\alpha\)} ;
    \vertex(f) at (2.5,1.5) {\(A_\beta\)};
    \diagram* {
       (a) -- [scalar, momentum'={[arrow shorten=0.15,xshift=-0.1cm]\(p_1\)}] (b) -- [graviton] (c) --[draw=none, momentum'={[arrow shorten=0.15,yshift=0.1cm]\(q\)}] (b),
       (b) --[scalar] (d),
       (d) --[photon, momentum'={[arrow shorten=0.15,xshift=0.1cm]\(p_2\)}] (e),
       (d) --[photon, momentum'={[arrow shorten=0.15,yshift=0.1cm]\(p_3\)}] (f);
    };
\end{feynman}
\end{tikzpicture}\qquad \quad
\begin{tikzpicture}[scale=0.9,baseline={-0.025cm*height("$=$")}]
  \begin{feynman}
    \vertex (a) at (-1.5, 1.5) {\(h_{\mu\nu}\)}; %
    \vertex (b) at (0,  0.52) ;                   %
    \vertex (c) at (1.5, 1.5) {\(A_{\beta}\)};    %
    \vertex (d) at (0, -0.52);                    %
    \vertex (e) at (-1.5,-1.5) {\(\phi\)} ;       %
    \vertex (f) at (1.5,-1.5) {\(A_\alpha\)};     %
    \diagram* {
       (a) -- [graviton] (b) -- [photon, momentum'={[arrow shorten=0.15,yshift=0.1cm]\(p_3\) }] (c),
       (a) -- [draw=none, momentum'={[arrow shorten=0.15,yshift=0.05cm]\(q\)}] (b),
       (b) -- [photon] (d),
       (e) -- [scalar, momentum'={[arrow shorten=0.15,xshift=-0.15cm]\(p_1\) }] (d),
       (d) -- [photon, momentum={[arrow shorten=0.15,yshift=-0.05cm,xshift=0.1cm]\(p_2\) }] (f);
    };
  \end{feynman}
\end{tikzpicture}
\qquad\quad
\begin{tikzpicture}[scale=0.9,baseline={-0.025cm*height("$=$")}]
  \begin{feynman}
    \vertex (a) at (-1.5, 1.5) {\(h_{\mu\nu}\)}; 
    \vertex (b) at (0,  0.52) ;                    
    \vertex (c) at (2,-1.5)  {\(A_{\alpha}\)} ;    %
    \vertex (d) at (0,-0.52);                      
    \vertex (e) at (-1.5,-1.5) {\(\phi\)}  ;       
    \vertex (f) at (2, 1.5)  {\(A_\beta\)};        %
    \diagram* {
       (a) -- [graviton] (b)
           -- [photon, momentum'={[arrow shorten=0.3, xshift=0.5cm, yshift=-0.42cm] \(p_2\)}] (c),
       (a) -- [draw=none, momentum'={[arrow shorten=0.15, yshift=0.05cm] \(q\)}] (b),
       (b) -- [photon] (d),
       (e) -- [scalar, momentum'={[arrow shorten=0.15, yshift=0.05cm] \(p_1\)}] (d),
       (d) -- [photon, momentum'={[arrow shorten=0.3, xshift=0.35cm, yshift=0.42cm] \(p_3\)}] (f);
    };
  \end{feynman}
\end{tikzpicture}
\label{fig:feynmandiagrams}
\caption{Feynman diagrams for graviton absorption-induced axion-decay. In the transverse-traceless (TT) gauge, the first diagram does not contribute, but is otherwise required to satisfy the Ward identity in the sum of diagrams. Because of the topological nature of the axion-photon coupling, an axion-photon-photon-graviton four-point interaction is absent.  }
\end{figure*}
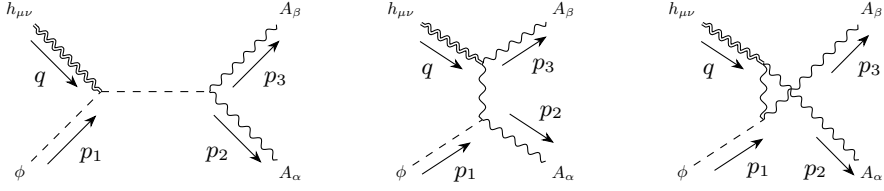

The matrix element $\M_{\phi+g\rightarrow\gamma\gamma}$ for graviton absorption-induced axion decay is found by summing the three Feynman diagrams shown in Fig.~\ref{fig:feynmandiagrams}. 
With the incoming graviton line flipped to an outgoing graviton line, the associated diagrams represent the emission processes contributing to the matrix element $\M_{\phi\rightarrow\gamma\gamma+g}$ instead. For the time evolution of the $\phi$ particle occupation number, we then have
\begin{widetext}
\begin{align}
   & \frac{d f_\phi}{\d t} =\frac{1}{4 E_{\vecp_1}} \int\prod_{i=2,3} \frac{d^3\vecp_i}{(2\pi)^3 2 E_{\vecp_i}}(2\pi)^4\delta^4(p_1-p_2-p_3)|\M_{\phi\rightarrow\gamma\gamma}|^2
    \left\{(1 + f_\phi)f_\gamma f_\gamma- f_\phi (1+f_\gamma)(1+f_\gamma)\right\}\notag\\
    &+\frac{1}{4 E_{\vecp_1}} \int \frac{d^3\vecq}{(2\pi)^3 2 \omega} \int\prod_{i=2,3} \frac{d^3\vecp_i}{(2\pi)^3 2 E_{\vecp_i}}(2\pi)^4\delta^4(p_1+q-p_2-p_3) |\M_{\phi+g\rightarrow\gamma\gamma}|^2
    \left\{(1 + f_\phi)(1+f_g)f_\gamma f_\gamma- f_\phi f_g (1+f_\gamma)(1+f_\gamma)\right\}\notag\\
    &+\frac{1}{4 E_{\vecp_1}} \int \frac{d^3\vecq}{(2\pi)^3 2 \omega }\int\prod_{i=2,3} \frac{d^3\vecp_i}{(2\pi)^3 2 E_{\vecp_i}}(2\pi)^4\delta^4(p_1-q-p_2-p_3) |\M_{\phi\rightarrow\gamma\gamma+g}|^2
    \left\{(1 + f_\phi)f_gf_\gamma f_\gamma- f_\phi (1+f_g) (1+f_\gamma)(1+f_\gamma)\right\}\,,
\end{align}
\end{widetext}
where the statistical factors $f_\phi$, $f_{\gamma}$, and $f_g$  arise, as usual, from commuting the field operators in the interaction Hamiltonian with the creation and annihilation operators of the initial and final states. The photon-polarization summed  squared vacuum decay matrix element for $\phi\to \gamma \gamma$ is given by $ |\M_{\phi\rightarrow\gamma\gamma}|^2  =g^2_{\phi\gamma} m_\phi^4/2$. For the graviton-induced processes $\phi+g\to \gamma \gamma $ and $\phi\to \gamma \gamma + g$, squaring and summing over both photon and graviton polarizations yields, respectively, 
\begin{subequations}
\begin{align}
|\M_{\phi+g\rightarrow \gamma\gamma}|^2 & =\frac{\kappa^2 g^2_{\phi\gamma}(m_\phi^8+s^4)}{8(s-m_\phi^2)^2}\,, \\
|\M_{\phi\rightarrow\gamma\gamma+g}|^2& =\frac{\kappa^2 g^2_{\phi\gamma}(m^8_\phi+\Bar{s}^4)}{8(\Bar{s}-m^2_\phi)^2}\,,
\end{align}
\end{subequations}
with $s\equiv (p_1+q)^2=(p_2+p_3)^2$, and $\Bar{s}\equiv (p_1-q)^2=(p_2+p_3)^2$. For $\omega\equiv q^0 \ll m_\phi$, which is the regime where the graviton occupation number becomes large, $|\M_{\phi+g\rightarrow \gamma\gamma}|^2\approx |\M_{\phi\rightarrow\gamma\gamma+g}|^2\approx \kappa^2 g^2_{\phi\gamma}m^6_\phi/(16 \omega^2)$. In this low-frequency regime, one can ignore the graviton four-momentum $q$ in the four-dimensional energy-momentum conserving Dirac delta functions. Assuming an isotropic occupation of gravitons, we see that absorption and emission of a graviton bring a relative enhancement factor as%
\begin{subequations}
\label{eq:appendixenhancements}
\begin{align}
     \text{absorption:}\ &\int \frac{\d^3\vecq}{(2\pi)^3 2 \omega} \frac{|\M_{\phi+g\rightarrow\gamma\gamma}|^2}{|\M_{\phi\rightarrow\gamma\gamma}|^2} f_g(\omega)\notag\\
     &\to \int_{\omega_{\rm min}}^{\Lambda} \frac{\d\omega}{\omega}  \frac{\kappa^2 m_\phi^2}{32\pi^2 }  f_g(\omega)\,, \\ \text{emission:}\ &\int \frac{\d^3\vecq}{(2\pi)^3 2 \omega} \frac{|\M_{\phi\rightarrow\gamma\gamma+g}|^2}{|\M_{\phi\rightarrow\gamma\gamma}|^2} [1+f_g(\omega)]\notag\\
     &\to \int_{\omega_{\rm min}}^\Lambda \frac{\d\omega}{\omega}  \frac{\kappa^2 m_\phi^2}{32\pi^2 }  [1+f_g(\omega)]\,,
 \end{align}
 \end{subequations}
where $\Lambda$ is an energy scale that defines a soft graviton, $\omega <\Lambda$. Here, $\omega_{\rm min}$ is an infrared energy cutoff. The integrals are very sensitive to $\omega_{\rm min}$. In particular, the emission enhancement factor diverges for $\omega_{\rm min}\rightarrow 0$ even if $f_g=0$, signalling the break-down of the leading-order analysis used here.

\subsection{Vacuum soft-theorem approach} As argued in the main text, we may get the above result and hence~\eqref{eq:enhancedDecay}  by using the soft-graviton theorem. Specializing to the case of axion decay, what is left to do is to evaluate the $B$-factor~\eqref{eq:Bfactor} and make contact with the Feynman-diagrammatic calculation of the previous section.
Because the final state photons (index $m,n=\{2,3\}$) are massless, one needs to take the limit $m_2\rightarrow 0$, $m_3\rightarrow 0$ in~\eqref{eq:Bfactor} carefully.
For $m=n \in\{2,3\}$, $\beta_{nn}\rightarrow 0$, and we obtain 
\begin{align}
    \frac{1}{\beta_{nn}}\ln\left(\frac{1+\beta_{nn}}{1-\beta_{nn}}\right)\rightarrow 2\, \quad (n=2,3) .
\end{align}
The corresponding contribution in $B$ hence vanishes because of the factor of $m_m m_n$ in front of the logarithm, cf.~\eqref{eq:Bfactor}.
For $m\neq n$ in the limit of the $n=2$ particle being massless we find
\begin{align}
    \beta_{2m}\simeq 1-\frac{m_2^2 m_m^2}{2(p_2\cdot p_m)^2} \qquad (m\neq 2)\,,
\end{align}
and similarly for $\beta_{3m}$.
The total $B$-factor is then given by
\begin{align}
    B=&\frac{\kappa^2}{32\pi^2}  \sum_{n,m;n\neq m}\eta_n \eta_m (p_n\cdot p_m) \ln\left(\frac{4(p_n\cdot p_m)^2}{m_n^2 m_m^2}\right) \notag \\ & + \frac{\kappa^2}{32\pi^2}  m_1^2 
    = \frac{\kappa^2 m_1^2}{32\pi^2} = \frac{\kappa^2 m_\phi^2}{32\pi^2}\,,
\end{align}
where in the last step, we have used $p_1\cdot p_2=p_2\cdot p_3=p_1\cdot p_3=m_1^2/2$ so that the first term vanishes. We may now identify the factors $\kappa^2 m_\phi^2/(32\pi^2)$ in~\eqref{eq:appendixenhancements} with~$B$ and obtain the relation
\begin{align}
    \int \mathrm{d}^2 \Omega_{\mathbf{q}}|\mathcal{M}_{\phi + g\to \gamma \gamma }|^2=\frac{2 (2 \pi)^3 B}{\omega^2}|\mathcal{M}_{\phi \rightarrow \gamma \gamma}|^2 ,
\end{align}
alluded to in the main text; the equivalent relation holds for $|\mathcal{M}_{\phi \to \gamma \gamma + g }|^2$.

\section{Mean emission and absorption spectra}

Here we derive the mean number of emitted and absorbed gravitons into the frequency bin $\Delta(\omega) \equiv [\omega, \omega + d\omega]$ accompanying a hard collision $\alpha \to \beta$. 
We denote the integrals of intensities defined in the main text 
by $M_{\rm em/abs}\equiv\int_\lambda^\Lambda \d\omega\,\mu_{\rm em/abs}(\omega)$. 
Since
$M_{\rm em}+M_{\rm abs}=B\ln(\Lambda/\lambda)+2B\int_\lambda^\Lambda\frac{\d\omega}{\omega}f_g(\omega)$,
the all-orders virtual factor in~\eqref{eq:Gamma} equals $\e^{-M_{\rm em}}\e^{-M_{\rm abs}}$, and~\eqref{eq:dGammaOmegas} takes the form
\begin{align}
\label{eq:poisson}
    d\Gamma^{(g)}_{\alpha\rightarrow\beta}
    =d\Gamma^0_{\alpha\rightarrow\beta}\,
    &\frac{\e^{-M_{\rm em}}}{N_{\rm em}!}\prod_{i=1}^{N_{\rm em}}\mu_{\rm em}(\omega_i)\,\d\omega_i
    \notag\\
    \times\,&\frac{\e^{-M_{\rm abs}}}{N_{\rm abs}!}\prod_{j=1}^{N_{\rm abs}}\mu_{\rm abs}(\omega_j)\,\d\omega_j\,.
\end{align}
The two factors give the probability for a Poisson process with respective multiplicities $N_{\rm em}$ and $N_{\rm abs}$ on the interval $[\lambda,\Lambda]$. The virtual factors provide the  \emph{normalization}: summing over multiplicity and integrating every frequency over the full range yields $\e^{-M}\sum_N M^N/N!=1$. 

The mean number of emitted gravitons in the frequency interval $\Delta(\omega)$ is obtained by weighting the count of quanta in~$\Delta(\omega)$ with the Poisson probability for emitting the unrestricted set $\{\omega_i \in [\lambda, \Lambda] \}_{i=1}^{N_{\rm em}}$ that multiplies the hard rate in~\eqref{eq:poisson}, integrating over all frequencies, and summing over the number of emitted quanta. Carrying out the frequency integrations yields
\begin{align}
\label{eq:mean}
    \langle N_{\Delta(\omega)}\rangle
    &=\sum_{N=1}^{\infty}N\,\frac{\e^{-M_{\rm em}}}{N!}\,
      M_{\rm em}^{N-1}\,\mu_{\rm em}(\omega)\,\d\omega
    =\mu_{\rm em}(\omega)\,\d\omega\,.
\end{align}
The factor $N$ counts, which of the distributed quanta falls into~$\Delta(\omega)$; an identical result holds for absorption with $\mu_{\rm abs}$. This is the basis of~\eqref{eq:expected-numbers}: the emission and absorption spectra are given, to all orders in~$G$ in the soft limit, by the leading-order intensities. The dependence on $\lambda $ and $\Lambda$ has cancelled between the virtual corrections and the sum over unobserved quanta in~\eqref{eq:mean}.

\bibliographystyle{apsrev4-1}
\bibliography{refs}

\end{document}